\documentclass[prd,amsmath,amssymb,superscriptaddress,preprintnumbers,twocolumn,nofootinbib,10pt]{revtex4-1}
\usepackage{graphicx}
\usepackage{dcolumn}
\usepackage{bm}
\usepackage{amssymb}
\usepackage{latexsym}
\usepackage{booktabs}
\usepackage{amsmath}
\usepackage{multirow}
\usepackage{url}
\usepackage{footnote}
\usepackage{float}
\usepackage{threeparttable}
\usepackage[colorlinks=true, linkcolor=blue, citecolor=blue]{hyperref}
\usepackage[bottom]{footmisc}

\usepackage[normalem]{ulem}
\usepackage{color}
\usepackage{array}
\usepackage{enumerate}
\usepackage{adjustbox}

\usepackage{makecell}
\usepackage{diagbox}
\usepackage{epstopdf}
\usepackage{epsfig}
\usepackage{longtable}
\usepackage{supertabular}
\usepackage{algorithm}
\usepackage{pifont}
\usepackage{algorithmic}
\usepackage{changepage}
\usepackage{setspace}
\begin{document}


\title{Robust evidence for dynamical dark energy in light of DESI DR2 and joint ACT, SPT, and Planck data}

\author{Tian-Nuo Li}\email{litiannuo@stumail.neu.edu.cn}
\affiliation{Liaoning Key Laboratory of Cosmology and Astrophysics, College of Sciences, Northeastern University, Shenyang 110819, China}

\author{Guo-Hong Du}
\affiliation{Liaoning Key Laboratory of Cosmology and Astrophysics, College of Sciences, Northeastern University, Shenyang 110819, China}

\author{Sheng-Han Zhou}
\affiliation{Liaoning Key Laboratory of Cosmology and Astrophysics, College of Sciences, Northeastern University, Shenyang 110819, China}

\author{Yun-He Li}
\affiliation{Liaoning Key Laboratory of Cosmology and Astrophysics, College of Sciences, Northeastern University, Shenyang 110819, China}

\author{Jing-Fei Zhang}
\affiliation{Liaoning Key Laboratory of Cosmology and Astrophysics, College of Sciences, Northeastern University, Shenyang 110819, China}

\author{Xin Zhang}\thanks{Corresponding author}\email{zhangxin@neu.edu.cn}
\affiliation{Liaoning Key Laboratory of Cosmology and Astrophysics, College of Sciences, Northeastern University, Shenyang 110819, China}
\affiliation{MOE Key Laboratory of Data Analytics and Optimization for Smart Industry, Northeastern University, Shenyang 110819, China}
\affiliation{National Frontiers Science Center for Industrial Intelligence and Systems Optimization, Northeastern University, Shenyang 110819, China}

\begin{abstract}

Recent baryon acoustic oscillation (BAO) measurements released by DESI, when combined with cosmic microwave background (CMB) data and type Ia supernova (SN) data, suggest a significant preference for dynamical dark energy (DDE) that exhibits the phantom-like behavior in the past and has transitioned into quintessence-like behavior today. In this work, we conduct a comprehensive analysis of six representative DDE parametrization models by utilizing the latest and most precise CMB data jointly from ACT, SPT, and Planck, in conjunction with BAO data from DESI DR2 and SN data from DESY5, PantheonPlus, and Union3. Our overall analysis indicates that the preference for DDE in the Quintom-B regime remains robust, regardless of the DDE parameterization model and the data combination employed. The trend of this preference is significantly strengthened with the support of DESY5 SN data. Specifically, when using the CMB+DESI+DESY5 data, for the Barboza-Alcaniz (BA) model, we obtain $w_0 = -0.785 \pm 0.047$ and $w_a = -0.43^{+0.10}_{-0.09}$, which significantly deviate from the $\Lambda$CDM values and provide evidence for DDE at the $4.2\sigma$ level. By the reconstruction of the dark energy equation of state $w(z)$, normalized dark energy density $f_{\mathrm{DE}}(z)$, and the deceleration parameter $q(z)$, we also observe clear departures from $\Lambda$CDM, further reinforcing the case for DDE. Furthermore, the Bayesian evidence analysis indicates that the Chevallier-Polarski-Linder, BA, and Exponential models are moderately favored relative to $\Lambda$CDM based on the CMB+DESI+DESY5 data. 
 
\end{abstract}
\maketitle

\section{Introduction}

The nature of dark energy (DE) remains one of the most profound unsolved problems in modern cosmology and fundamental physics. DE was introduced to explain the accelerated expansion of the universe, a phenomenon first observed through type Ia supernova (SN) in 1998 \cite{SupernovaSearchTeam:1998fmf,SupernovaCosmologyProject:1998vns} and later confirmed by observations of cosmic microwave background (CMB) \cite{WMAP:2003elm} and baryon acoustic oscillation (BAO) \cite{SDSS:2005xqv}. These cosmological data seem to support the existence of the simplest theoretical interpretation, the cosmological constant $\Lambda$, with the DE equation of state (EoS) parameter $w = -1$, which is a key element of the standard $\Lambda$ Cold Dark Matter ($\Lambda$CDM) model. Although the $\Lambda$CDM model has been successful in explaining observational data, it still faces some issues, including the ``fine-tuning'' and ``cosmic coincidence'' problems~\cite{Sahni:1999gb,Bean:2005ru}, as well as emerging observational tensions in measurements, such as the $H_0$ tension \cite{Verde:2019ivm,Vagnozzi:2019ezj,DiValentino:2020zio,DiValentino:2021izs,Vagnozzi:2021gjh,Vagnozzi:2021tjv,DiValentino:2022fjm,Vagnozzi:2023nrq} and the $S_8$ tension \cite{DiValentino:2020vvd}. The completeness of the $\Lambda$CDM model has been challenged by these issues, further motivating the exploration of alternative cosmological scenarios \cite{Boisseau:2000pr,Chevallier:2000qy,Li:2004rb,Zhang:2005hs, Zhang:2005yz,Zhang:2007sh,Ma:2007av,Zhang:2009un,Zhang:2014ifa,Li:2014eha,Cai:2015emx,Wang:2016lxa,Feng:2017usu,Poulin:2018cxd,Pan:2019gop,Pan:2020zza,Wang:2021kxc,Colgain:2021pmf,Yin:2023srb,Yao:2023qve,Yao:2023ybs,Lyu:2025nsd,MV:2025yjt}; for a recent comprehensive review, see Ref.~\cite{CosmoVerseNetwork:2025alb}.

In particular, since the beginning of 2024, the $\Lambda$CDM model has been more severely challenged by new BAO measurements from the first data release of the Dark Energy Spectroscopic Instrument (DESI), with the recent second data release (DR2) data intensifying this challenge~\cite{DESI:2024mwx,DESI:2025zgx}. When DESI DR2 BAO data are combined with the CMB data from Planck and the Atacama Cosmology Telescope (ACT), and SN datasets (whether from PantheonPlus, Union3, or DESY5), significant deviations from the $\Lambda$CDM paradigm are revealed, strongly favoring dynamical DE (DDE). Specifically, within the Chevallier-Polarski-Linder (CPL) parameterization of the DE EoS, $w(a) = w_0 + w_a(1 - a)$~\cite{Chevallier:2000qy,Linder:2002et}, where $w_0$ represents the current DE EoS and $w_a$ quantifies its dynamical evolution, we observe a preference for DDE\footnote{The EoS parameter of DE evolves across $w = -1$ (transitioning from a past phantom-like behavior with $w < -1$ to a present quintessence-like behavior with $w > -1$), a phenomenon known as the Quintom scenario \cite{Feng:2004ad,Zhang:2005kj}.} with $w_0 > -1$ and $w_a < 0$, at a statistical significance of $2.8\sigma$ to $4.2\sigma$ depending on the specific SN data used. These findings have sparked extensive debates on the nature of DE, including scalar field DE models \cite{Li:2024qus,Fazzari:2025lzd,Toomey:2025xyo,Goh:2025upc,Wolf:2025acj,Luciano:2025ykr,Plaza:2025nip,Moretti:2025gbp,Li:2025vqt,Wu:2025vfs}, interacting DE models \cite{Sabogal:2024yha,Li:2025owk,Li:2025ula,Goswami:2025uih,You:2025uon,Li:2024qso,Pan:2025qwy,Giare:2024smz,Hussain:2025uye,Yang:2025uyv,Wang:2025znm,Shah:2025ayl,Silva:2025hxw,Samanta:2025oqz,Zhu:2025lrk,Li:2025muv,Zhang:2025dwu,vanderWesthuizen:2025rip,Li:2026xaz}, early DE models \cite{Wang:2024dka,Yashiki:2025loj,Pang:2025lvh}, phenomenological DE models \cite{Giare:2024gpk,Wu:2025wyk,Li:2025ops,Barua:2025ypw,Ling:2025lmw,Alam:2025epg,Huang:2025som,Qiang:2025cxp,Cheng:2025lod}, and other aspects of cosmological physics \cite{Chen:2025wwn,Araya:2025rqz,Li:2025eqh,Li:2025dwz,Paliathanasis:2025xxm,Du:2024pai,Jiang:2024viw,Du:2025iow,Feng:2025mlo,RoyChoudhury:2025dhe,Du:2025xes,Zhou:2025nkb,Cai:2025mas,Li:2025htp,Jia:2025poj,RoyChoudhury:2025iis,Pedrotti:2025ccw,Paliathanasis:2025kmg,Wang:2025vtw,Liu:2025myr,Yao:2025kuz,Adam:2025kve,Yang:2025oax,Afroz:2025iwo,Liu:2025evk,Paul:2025wix,Braglia:2025gdo,Guin:2025xki,Efstratiou:2025iqi,Luciano:2025dhb,Akarsu:2025nns,Yao:2025twv,Zhou:2025dxo}.

Undoubtedly, the indications of DDE reported by DESI represent one of the most compelling signs of new physics beyond the $\Lambda$CDM model, with significant implications for our understanding of the universe. Therefore, given the importance of this result, numerous studies have been conducted to further test the robustness of these findings, including potential systematic issues in these datasets \cite{Colgain:2024xqj,Naredo-Tuero:2024sgf,Efstathiou:2024xcq,Colgain:2024ksa}, various parametrizations of the DE EoS \cite{Giare:2024gpk,Arora:2025msq}, and different combinations of datasets \cite{Giare:2024oil,Park:2024vrw,Reeves:2025axp,Liu:2025mub,Song:2025bio,Ishak:2025cay}. For example, potential systematic effects in DESI BAO measurements may have contributed to the reported preference for DDE. Specifically, the DESI BAO measurements at $z = 0.51$ and $z = 0.71$ show approximately $2\sigma$ and $3\sigma$ tensions with predictions based on the Planck best-fit $\Lambda$CDM model \cite{Colgain:2024xqj,Naredo-Tuero:2024sgf}. Additionally, the evidence for DDE may also primarily arise from systematic effects in the DESY5 SN data, as the parameter range favored by the uncorrected DESY5 sample deviates from many other cosmological datasets \cite{Efstathiou:2024xcq, Colgain:2024ksa, Notari:2024zmi, Huang:2024qno}. Interestingly, \citet{Giare:2024gpk} found that the Barboza-Alcaniz (BA) parameterization provides a better fit than the CPL parameterization when combining CMB, DESI BAO, and SN data. From data-combination perspective, Ref.~\cite{Giare:2025pzu} tests the robustness of the DDE evidence using 35 different dataset combinations under the CPL model, showing that combinations including both PantheonPlus SN and Sloan Digital Sky Survey BAO significantly weaken the preference for DDE. Therefore, it is also worth investigating the impact of different combinations of observational datasets, as these diverse data are likely to effectively break parameter degeneracies, thereby improving the measurement of DE EoS parameters.

In recent months, the ground-based CMB experiments, the ACT \cite{AtacamaCosmologyTelescope:2025blo} and the South Pole Telescope (SPT) \cite{SPT-3G:2025bzu} have released updated measurements of temperature and polarization anisotropies, extending the observational reach to higher multipoles (i.e., smaller angular scales). These datasets constitute the most precise measurements of small-scale CMB polarization to date. In particular, when combined with Planck’s large-scale CMB data, they yield substantially tighter constraints on cosmological parameters. Recently, the ACT and SPT data have been widely used in cosmological studies ~\cite{Peng:2025tqt, Peng:2025vda, Pang:2024wul, Giare:2024oil, AtacamaCosmologyTelescope:2025nti, Poulin:2025nfb, SPT-3G:2025vyw, Yin:2025fmj, Wang:2025djw}. For example, \citet{Peng:2025vda} utilize the latest ACT and SPT CMB data combined with Planck CMB data to constrain the periodic oscillations in the primordial power spectrum, providing state-of-the-art CMB constraints on primordial oscillations. Specifically, Ref.~\cite{ACT:2025qjh} combines CMB lensing measurements from ACT, SPT, and Planck (providing the most stringent CMB lensing constraints to date), along with DESI DR2, which improves the constraint on the amplitude of matter fluctuations to $\sigma_8 = 0.829 \pm 0.009$ (a 1.1\% determination). These increasingly precise CMB, BAO, and SN datasets have ushered in a new era of precision cosmology, enabling precise tests of physics both within and beyond the standard $\Lambda$CDM framework.

Given these critical issues, it is essential to use the most precise current dataset combinations to test various DDE parameterizations. In this work, we use the most precise datasets, including CMB data from ACT, SPT, and Planck, along with BAO data from DESI DR2 and SN data from PantheonPlus, DESY5 and Union3, to constrain various DDE models. We consider six representative DDE parameterizations to conduct a comprehensive and robust analysis, including CPL, Jassal-Bagla-Padmanabhan (JBP), BA, Exponential (EXP), Logarithmic (LOG), and Sinusoidal (SIN) parameterization models. In addition, we investigate the evolution of DE EoS in these models and evaluate the evidence for DDE using the current observational data. 

This work is organized as follows. In Sec.~\ref{sec2}, we briefly introduce the DDE models considered in this work, along with the cosmological data utilized in the analysis. In Sec.~\ref{sec3}, we report the constraint results and make some relevant discussions. The conclusion is given in Sec.~\ref{sec4}.

\section{Methodology and data}\label{sec2}

\subsection{Brief description of the DDE models}\label{sec2.1}

In the flat Friedmann-Lema\^{\i}tre-Robertson-Walker metric, the dimensionless Hubble parameter $E(a)$ can be written as
\begin{align}
E^2(a)&\equiv\frac{H^2(a)}{H_0^2}=(\Omega_\mathrm{b}+\Omega_\mathrm{c})a^{-3}+\Omega_\mathrm{r} a^{-4}+\Omega_\mathrm{DE}f_\mathrm{DE}(a),
\end{align}
where $H(a)$ is the Hubble parameter, $a$ is the scale factor, $\Omega_\mathrm{b}$, $\Omega_\mathrm{c}$, $\Omega_\mathrm{r}$, and $\Omega_\mathrm{DE}$ are the present-time energy density parameters in baryons, cold dark matter, radiation, and DE, respectively. $f_\mathrm{DE}(z)$ represents the normalized $a$-dependent density of DE by its present value, given by
\begin{equation}\label{eq2}
f_\mathrm{DE}(a) = \exp\left(3\int_{a}^{1} \frac{1 + w(a^\prime)}{a^\prime} \mathrm{d}a^\prime \right).
\end{equation}
Here, $w(a)=P_\mathrm{DE}(a)/\rho_\mathrm{DE}(a)$ is the EoS of DE, where $P_\mathrm{DE}(a)$ and $\rho_\mathrm{DE}(a)$ are the pressure and energy density of DE.

In this work, we examine six different parametrizations for the EoS of DE, which are described as follows:

\begin{itemize}

\item \textbf{\texttt{Chevallier-Polarski-Linder Parametrization (CPL)}:}

The CPL parametrization, proposed by Chevallier, Polarski and Linder~\cite{Chevallier:2000qy,Linder:2002et}, is basic and widely used in DDE studies. The EoS of DE in this context is expressed as
\begin{equation}
w(a) = w_0 + w_a (1 - a), 
\end{equation}
where $w_0$ and $w_a$ are constants representing the present value of the EoS and its evolution with the scale factor. This parameterization is capable of accurately reconstructing various EoS for scalar fields and the resulting distance-redshift relations, providing a concise yet effective description of the dynamical evolution of DE.

\item \textbf{\texttt{Jassal-Bagla-Padmanabhan Parametrization (JBP)}:}

The JBP parameterization~\cite{Jassal:2005qc} describes the EoS of DE as
\begin{equation}
w(a) = w_0 + w_a a \left(1 - a \right).
\end{equation}
This parameterization includes both linear and quadratic terms in the scale factor and can lead to differences in the DE behavior at low redshift compared to CPL, especially when $a$ is close to 1.

\item \textbf{\texttt{Barboza-Alcaniz Parametrization (BA)}:}

The BA parameterization~\cite{Barboza:2008rh} is expressed as
\begin{equation}
w(a) = w_0 + w_a \frac{1 - a}{a^2 + (1 - a)^2}.
\end{equation}
This parameterization remains well-behaved at both low and high redshifts, showing linear behavior at low redshifts while allowing deviations from the baseline CPL model.

\item \textbf{\texttt{Exponential Parametrization (EXP)}:}

Unlike above three polynomial parameterizations, we next consider an exponential form for the EoS of DE~\cite{Dimakis:2016mip,Pan:2019brc}, given by
\begin{equation}
w(a) = (w_0 - w_a) + w_a \exp\left(1 - a \right).
\end{equation}
When $a$ is close to 1, this parameterization reduces to the CPL parameterization at the first order of its Taylor expansion. While $a$ deviates significantly from 1, it introduces (small) higher-order deviations from the linear CPL form \cite{Najafi:2024qzm}.

\item \textbf{\texttt{Logarithmic Parametrization (LOG)}:}

The CPL and similar polynomial parameterizations (such as JBP and BA which were mentioned earlier), face the issue that their EoS diverges in the far future. As a result, while these models can correctly describe the past evolution, they are unable to portray future evolution. To overcome this limitation, \citet{Ma:2011nc} proposed a logarithmic parameterization of EoS, expressed as
\begin{equation}
w(a) = w_0 - w_a \left[ a\ln\left(1+\frac{1}{a}\right) - \ln 2 \right].
\end{equation}
This parameterization successfully circumvents the future divergence problem inherent in the CPL parameterization, and can be utilized to investigate the dynamical characteristics of DE throughout the whole evolutionary history.

\item \textbf{\texttt{Sinusoidal Parametrization (SIN)}:}

Similar to the LOG parameterization, \citet{Ma:2011nc} also proposed a sinusoidally oscillating parameterization to circumvent the future divergence problem, whose EoS is given by
\begin{equation}
w(a) = w_0 - w_a \left[ a\sin \left(\frac{1}{a}\right) - \sin(1) \right].
\end{equation}

\end{itemize}

\subsection{Cosmological data}\label{sec2.2}

\begin{table}[t]
\caption{Flat priors on the main cosmological parameters constrained in this paper.}
\begin{center}
\renewcommand{\arraystretch}{1.8}
\begin{tabular}{@{\hspace{0.4cm}}c@{\hspace{0.6cm}} c@{\hspace{0.6cm}} c @{\hspace{0.4cm}} }
\hline\hline
\textbf{Model}       & \textbf{Parameter}       & \textbf{Prior}\\
\hline
$\Lambda$CDM        & $\Omega_{\rm b} h^2$                     & $\mathcal{U}$[0.005\,,\,0.1] \\
                    & $\Omega_{\rm c} h^2$                     & $\mathcal{U}$[0.01\,,\,0.99] \\
                    & $H_0$                                    & $\mathcal{U}$[20\,,\,100] \\
                    & $\tau$                                   & $\mathcal{U}$[0.01\,,\,0.8] \\
                    & $\log(10^{10}A_{\rm s})$                 & $\mathcal{U}$[1.61\,,\,3.91] \\
                    & $n_{\rm s}$                              & $\mathcal{U}$[0.8\,,\,1.2] \\
\hline
DDE                 & $w_0$                                  & $\mathcal{U}$[-3\,,\,2] \\
                    & $w_a$                                  & $\mathcal{U}$[-8\,,\,3] \\
\hline\hline
\end{tabular}
\label{tab1}
\end{center}	
\end{table}

Table~\ref{tab1} lists the free parameters of these models and the uniform priors applied. The parameter set for the $\Lambda$CDM model is $\bm{\theta}_{\Lambda\mathrm{CDM}}=\{\Omega_{\rm b} h^2$, $\Omega_{\rm c} h^2$, $H_0$, $\tau$, $\log(10^{10}A_{\rm s})$, $n_{\rm s}\}$. For the DDE models, the parameter sets are $\bm{\theta}_{\mathrm{DDE}}=\{\bm{\theta}_{\Lambda\mathrm{CDM}},w_0,w_a\}$. The theoretical models are computed using a modified version of the {\tt CAMB} code \cite{Lewis:1999bs}, which includes various DDE parameterizations. We employ the publicly available sampler {\tt Cobaya}\footnote{\url{https://github.com/CobayaSampler/cobaya}.} \cite{Torrado:2020dgo} to conduct a Markov Chain Monte Carlo (MCMC) analysis, assessing chain convergence with the Gelman-Rubin statistic, $R - 1 < 0.02$ \cite{Gelman:1992zz}. The resulting MCMC chains are then analyzed using the public package {\tt GetDist}\footnote{\url{https://github.com/cmbant/getdist}.} \cite{Lewis:2019xzd}. We place constraints on the models using the latest observational data, reporting the best-fit values and the $1\sigma$--$2\sigma$ confidence levels for the key parameters of \{$H_{0}$, $\Omega_{\mathrm{m}}$, $w_0$, $w_a$\}. 

The datasets used are as follows:
\begin{itemize}

\item \textbf{\texttt{CMB}:} We utilize the following combination of CMB data: (i) the large-scale temperature (TT, $2 \leq \ell \leq 30$) and E-mode polarization (EE, $2 \leq \ell \leq 30$) spectra from the \texttt{Commander} and \texttt{SimAll} likelihoods of Planck, as well as the small-scale TT ($30 \leq \ell \leq 1000$) and TE/EE ($30 \leq \ell \leq 600$) spectra from the \texttt{CamSpec} likelihood~\cite{Efstathiou:2019mdh,Rosenberg:2022sdy}; (ii) the \texttt{ACT-lite} likelihood for ACT DR6\footnote{\url{https://github.com/ACTCollaboration/DR6-ACT-lite}.}, providing the small-scale temperature and polarization ($1000 \leq \ell \leq 8500$ for TT, $600 \leq \ell \leq 8500$ for TE/EE) power spectra~\cite{AtacamaCosmologyTelescope:2025blo}; (iii) the SPT-3G Main field data\footnote{\url{https://github.com/SouthPoleTelescope/spt_candl_data}.}, offering the temperature and polarization (TT/TE/EE) anisotropy spectra from the \texttt{SPT-lite} likelihood~\cite{SPT-3G:2025bzu}; (iv) the CMB lensing power spectrum measurements from ACT and Planck lensing-reconstruction maps combined with SPT-3G lensing data\footnote{\url{https://github.com/qujia7/spt_act_likelihood}.}~\cite{Carron:2022eyg, ACT:2023dou, ACT:2023kun, ACT:2025qjh}. Note that to properly combine the Planck data with those from ACT and SPT-3G, we follow the methodology of the ACT collaboration \cite{AtacamaCosmologyTelescope:2025blo} by applying a multipole cut of $\ell < 1000$ to the Planck TT power spectrum and $\ell < 600$ to the TE/EE spectra. A Gaussian prior on the optical depth, $\tau = 0.0566 \pm 0.0058$, is adopted in our analysis.

\item \textbf{\texttt{DESI}:} The DESI DR2 BAO measurements utilized in this work are detailed in Table IV of Ref.~\cite{DESI:2025zgx}. These measurements include the transverse comoving distance $D_{\mathrm{M}}/r_{\mathrm{d}}$, the angle-averaged distance $D_{\mathrm{V}}/r_{\mathrm{d}}$, and the Hubble horizon $D_{\mathrm{H}}/r_{\mathrm{d}}$, where $r_{\mathrm{d}}$ denotes the comoving sound horizon at the drag epoch.

\item \textbf{\texttt{DESY5}:} The DESY5\footnote{\url{https://github.com/des-science/DES-SN5YR}.} sample comprises 1829 type Ia supernovae (SNe), combining 1635 photometrically classified SNe from the full 5-year data of the Dark Energy Survey collaboration, with the redshift range $0.1 < z < 1.3$, and 194 low-redshift SNe from complementary samples~\cite{Hicken:2009df,Hicken:2012zr,Krisciunas:2017yoe,Foley:2017zdq}, spanning the redshift range $0.025 < z < 0.1$ \cite{DES:2024jxu}.

\item \textbf{\texttt{PantheonPlus}:} The PantheonPlus\footnote{\url{https://github.com/PantheonPlusSH0ES/DataRelease}.} sample comprises 1550 spectroscopically confirmed SNe from 18 different surveys, spanning the redshift range $0.01 < z < 2.26$ \cite{Brout:2022vxf}.

\item \textbf{\texttt{Union3}:} The Union3\footnote{\url{https://github.com/rubind/union3_release}.} compilation has 2087 SNe and 1363 of which are common to PantheonPlus, covering the redshift range $0.01 < z < 2.26$ \cite{Rubin:2023jdq}.

\end{itemize}

\section{Results and discussions}\label{sec3}

\begin{table*}[htbp]
\centering
\caption{Fitting results ($1\sigma$ confidence level) in the $\Lambda$CDM, CPL, JBP, BA, EXP, LOG, and SIN models from the CMB+DESI, CMB+DESI+DESY5, CMB+DESI+PantheonPlus, and CMB+DESI+Union3 data. Here, $H_{0}$ is in units of ${\rm km}~{\rm s}^{-1}~{\rm Mpc}^{-1}$. Negative values of $\Delta \chi^2_\mathrm{MAP} \equiv \chi^2_{\rm DDE}-\chi^2_{\Lambda\rm CDM}$ indicate a better fit of DDE to the data compared to $\Lambda$CDM.}
\label{tab2}
\setlength{\tabcolsep}{2mm}
\renewcommand{\arraystretch}{1.3}
\small
\begin{tabular}{lc c c c c c}
\hline 
\hline
Model/Dataset & $H_0$ &$\Omega_{\mathrm{m}}$& $w_0$ & $w_a$ & $\Delta \chi^2_{\text{MAP}}$ & $N\sigma$ \\
\hline
\multicolumn{7}{l}{$\bm{\Lambda}$\textbf{CDM}} \\
CMB+DESI                & $68.12\pm 0.25$ & $0.3043\pm 0.0033$ & - & - & -  & - \\
CMB+DESI+DESY5          & $67.98\pm 0.24$ & $0.3062\pm 0.0032$ & - & - & -  & - \\
CMB+DESI+PantheonPlus   & $68.05\pm 0.24$ & $0.3052\pm 0.0033$ & - & - & -  & - \\
CMB+DESI+Union3         & $68.05\pm 0.24$ & $0.3052\pm 0.0033$ & - & - & -  & - \\
\hline
\multicolumn{7}{l}{$\bm{\textbf{CPL}}$} \\
CMB+DESI                & $63.80^{+1.70}_{-2.00}$  & $0.3510\pm 0.0210$       & $-0.430\pm 0.210$        & $-1.72\pm 0.57$            & $-11.2$ & $2.9\sigma$ \\
CMB+DESI+DESY5          & $66.86\pm 0.55$       & $0.3187\pm 0.0054$     & $-0.749\pm 0.057$      & $-0.88^{+0.23}_{-0.19}$    & $-18.9$  & $3.9\sigma$ \\
CMB+DESI+PantheonPlus   & $67.66\pm 0.59$       & $0.3110\pm 0.0056$     & $-0.833\pm 0.055$      & $-0.65\pm 0.20$            & $-7.9$  & $2.3\sigma$ \\
CMB+DESI+Union3         & $66.03\pm 0.82$       & $0.3270\pm 0.0083$     & $-0.664\pm 0.085$      & $-1.10^{+0.29}_{-0.26}$    & $-15.1$  & $3.5\sigma$ \\
\hline
\multicolumn{7}{l}{$\bm{\textbf{JBP}}$} \\
CMB+DESI                & $66.84^{+0.98}_{-2.00}$  & $0.3190^{+0.0190}_{-0.0100}$ & $-0.420^{+0.240}_{-0.095}$ & $-3.14^{+1.10}_{-1.40}$                & $-12.4$  & $3.1\sigma$ \\
CMB+DESI+DESY5          & $66.74\pm 0.56$         & $0.3192\pm 0.0055$       & $-0.649^{+0.081}_{-0.073}$ & $-1.99\pm 0.45$        & $-19.8$  & $4.1\sigma$ \\
CMB+DESI+PantheonPlus   & $67.71\pm 0.60$         & $0.3099\pm 0.0057$       & $-0.793\pm 0.076$         & $-1.29\pm 0.45$        & $-11.4$  & $2.9\sigma$ \\
CMB+DESI+Union3         & $65.89^{+0.82}_{-0.92}$ & $0.3278\pm 0.0089$       & $-0.530^{+0.130}_{-0.120}$ & $-2.58\pm0.63$               & $-13.5$  & $3.2\sigma$ \\
\hline
\multicolumn{7}{l}{$\bm{\textbf{BA}}$} \\
CMB+DESI                & $63.90\pm 1.90$         & $0.3500\pm 0.0210$       & $-0.500\pm 0.180$        & $-0.83^{+0.29}_{-0.25}$  & $-11.8$  & $3.0\sigma$ \\
CMB+DESI+DESY5          & $66.89\pm 0.56$       & $0.3185\pm 0.0055$     & $-0.785\pm 0.047$      & $-0.43^{+0.10}_{-0.09}$& $-21.2$  & $4.2\sigma$ \\
CMB+DESI+PantheonPlus   & $67.66\pm 0.59$       & $0.3110\pm 0.0056$     & $-0.856\pm 0.048$      & $-0.32\pm 0.09$        & $-15.2$  & $3.5\sigma$ \\
CMB+DESI+Union3         & $66.11\pm 0.82$       & $0.3262\pm 0.0083$     & $-0.714\pm 0.073$      & $-0.52^{+0.13}_{-0.12}$ & $-17.6$  & $3.8\sigma$ \\
\hline
\multicolumn{7}{l}{$\bm{\textbf{EXP}}$} \\
CMB+DESI                & $64.30\pm 1.70$         & $0.3450\pm 0.0190$       & $-0.520\pm 0.180$        & $-1.18^{+0.43}_{-0.38}$ & $-13.6$  & $3.3\sigma$ \\
CMB+DESI+DESY5          & $66.88\pm 0.56$       & $0.3186\pm 0.0055$     & $-0.772\pm 0.051$      & $-0.65^{+0.17}_{-0.15}$ & $-18.1$  & $3.9\sigma$ \\
CMB+DESI+PantheonPlus   & $67.65\pm 0.58$       & $0.3112\pm 0.0056$     & $-0.846\pm 0.049$      & $-0.49\pm 0.15$         & $-13.3$  & $3.2\sigma$ \\
CMB+DESI+Union3         & $66.14\pm 0.82$       & $0.3261\pm 0.0083$     & $-0.700\pm 0.079$      & $-0.80^{+0.22}_{-0.19}$ & $-14.3$  & $3.4\sigma$ \\
\hline
\multicolumn{7}{l}{$\bm{\textbf{LOG}}$} \\
CMB+DESI                & $64.80^{+1.60}_{-1.80}$  & $0.3410\pm 0.0180$       & $-0.590\pm 0.160$        & $-4.70^{+1.80}_{-1.50}$     & $-10.2$  & $2.7\sigma$ \\
CMB+DESI+DESY5          & $66.94\pm 0.56$       & $0.3183\pm 0.0055$     & $-0.788\pm 0.049$      & $-2.83^{+0.72}_{-0.65}$  & $-18.5$  & $3.9\sigma$ \\
CMB+DESI+PantheonPlus   & $67.66\pm 0.58$       & $0.3112\pm 0.0054$     & $-0.858^{+0.044}_{-0.050}$ & $-2.12^{+0.71}_{-0.61}$ & $-13.4$  & $3.2\sigma$ \\
CMB+DESI+Union3         & $66.20\pm 0.82$       & $0.3255\pm 0.0083$     & $-0.720^{+0.070}_{-0.078}$ & $-3.48^{+0.96}_{-0.85}$ & $-15.0$   & $3.5\sigma$ \\
\hline
\multicolumn{7}{l}{$\bm{\textbf{SIN}}$} \\
CMB+DESI                & $65.60\pm 1.60$         & $0.3320\pm 0.0170$       & $-0.700^{+0.120}_{-0.140}$ & $-1.45^{+0.67}_{-0.46}$ & $-10.3$  & $2.8\sigma$ \\
CMB+DESI+DESY5          & $67.01\pm 0.55$       & $0.3177\pm 0.0055$     & $-0.824\pm 0.041$       & $-0.98^{+0.27}_{-0.24}$ & $-17.9$  & $3.8\sigma$ \\
CMB+DESI+PantheonPlus   & $67.69\pm 0.59$       & $0.3111\pm 0.0056$     & $-0.877\pm 0.042$       & $-0.78^{+0.26}_{-0.21}$ & $-10.6$  & $2.8\sigma$ \\
CMB+DESI+Union3         & $66.39\pm 0.78$       & $0.3239\pm 0.0078$     & $-0.772\pm 0.063$       & $-1.18^{+0.36}_{-0.28}$ & $-16.7$  & $3.7\sigma$ \\
\hline
\hline
\end{tabular}
\end{table*}

\begin{figure*}[htpb]
\includegraphics[scale=0.4]{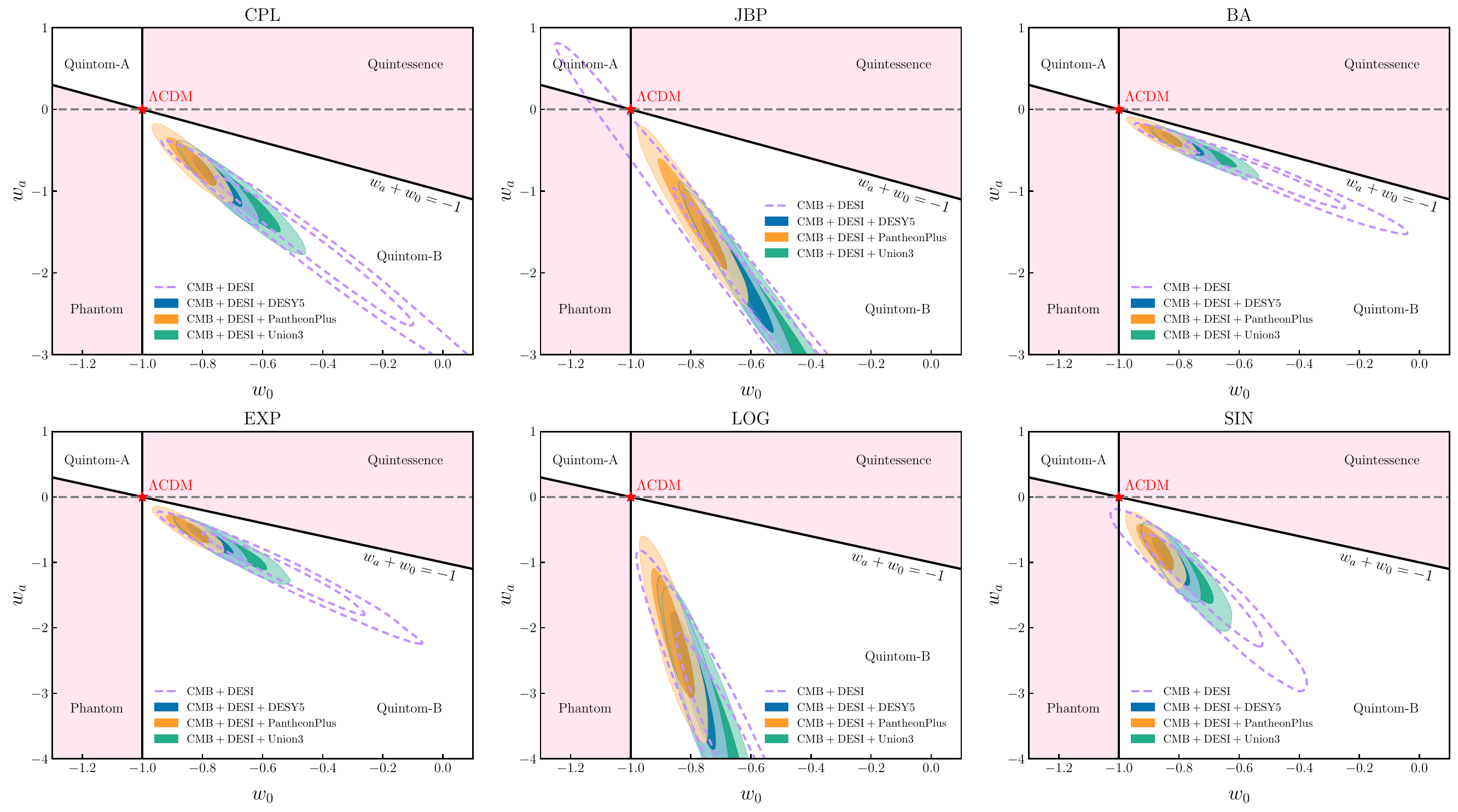}
\centering
\caption{\label{fig1} Two-dimensional marginalized contours ($1\sigma$ and $2\sigma$ confidence levels) in the $w_0$--$w_a$ plane by using the CMB+DESI, CMB+DESI+DESY5, CMB+DESI+PantheonPlus, and CMB+DESI+Union3 data in the CPL, JBP, BA, EXP, LOG, and SIN models. We have included boundary lines to delineate the regions of quintessence, phantom, and quintom. The red pentagram marks the $\Lambda$CDM model.}
\end{figure*}

\begin{figure}[htpb!]
\includegraphics[scale=0.55]{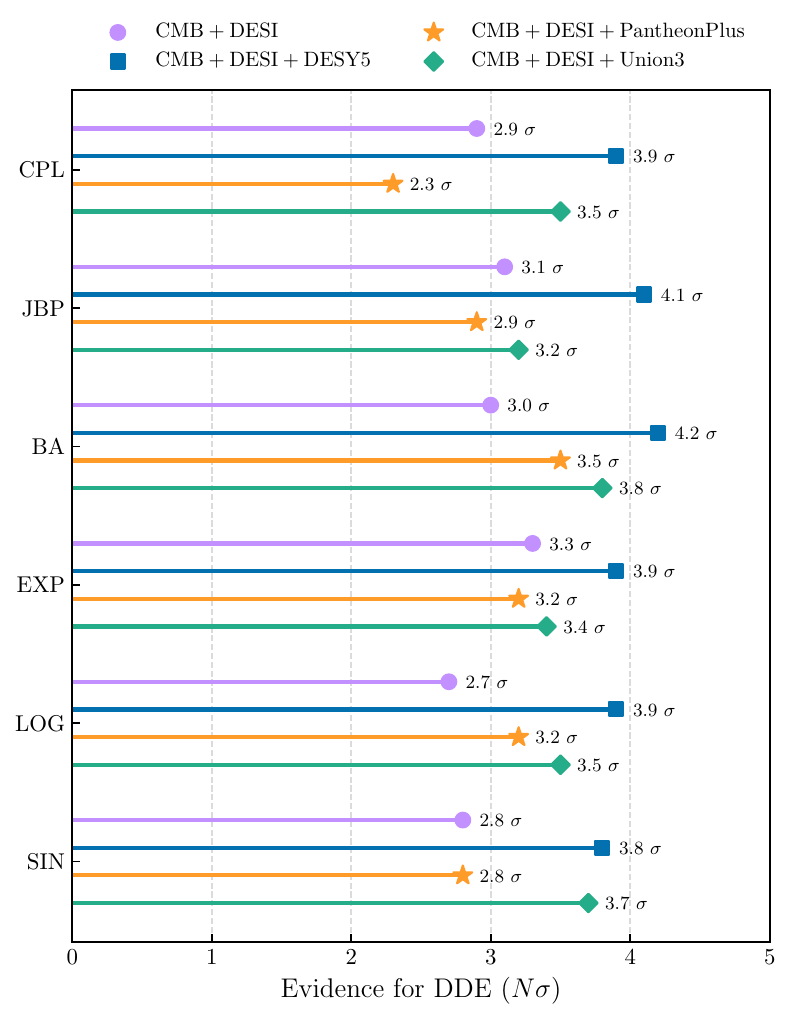}
\centering
\caption{\label{fig2} The evidence for DDE in the CPL, JBP, BA, EXP, LOG, and SIN models using CMB+DESI, CMB+DESI+DESY5, CMB+DESI+PantheonPlus, and CMB+DESI+Union3 data.}

\end{figure}

\begin{figure*}[htpb]
\centering
\includegraphics[width=.9\linewidth]{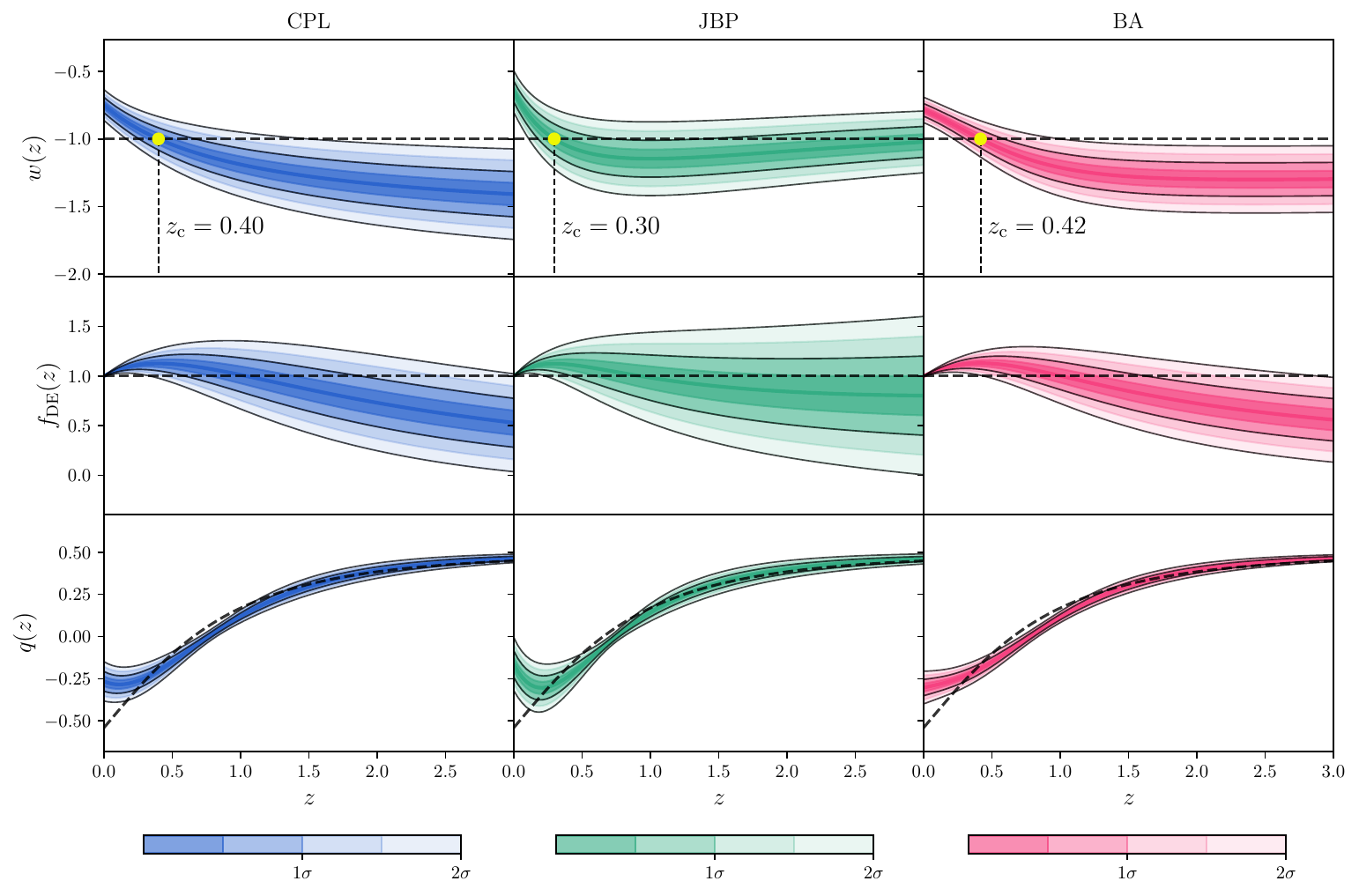}
\par\bigskip          
\includegraphics[width=.9\linewidth]{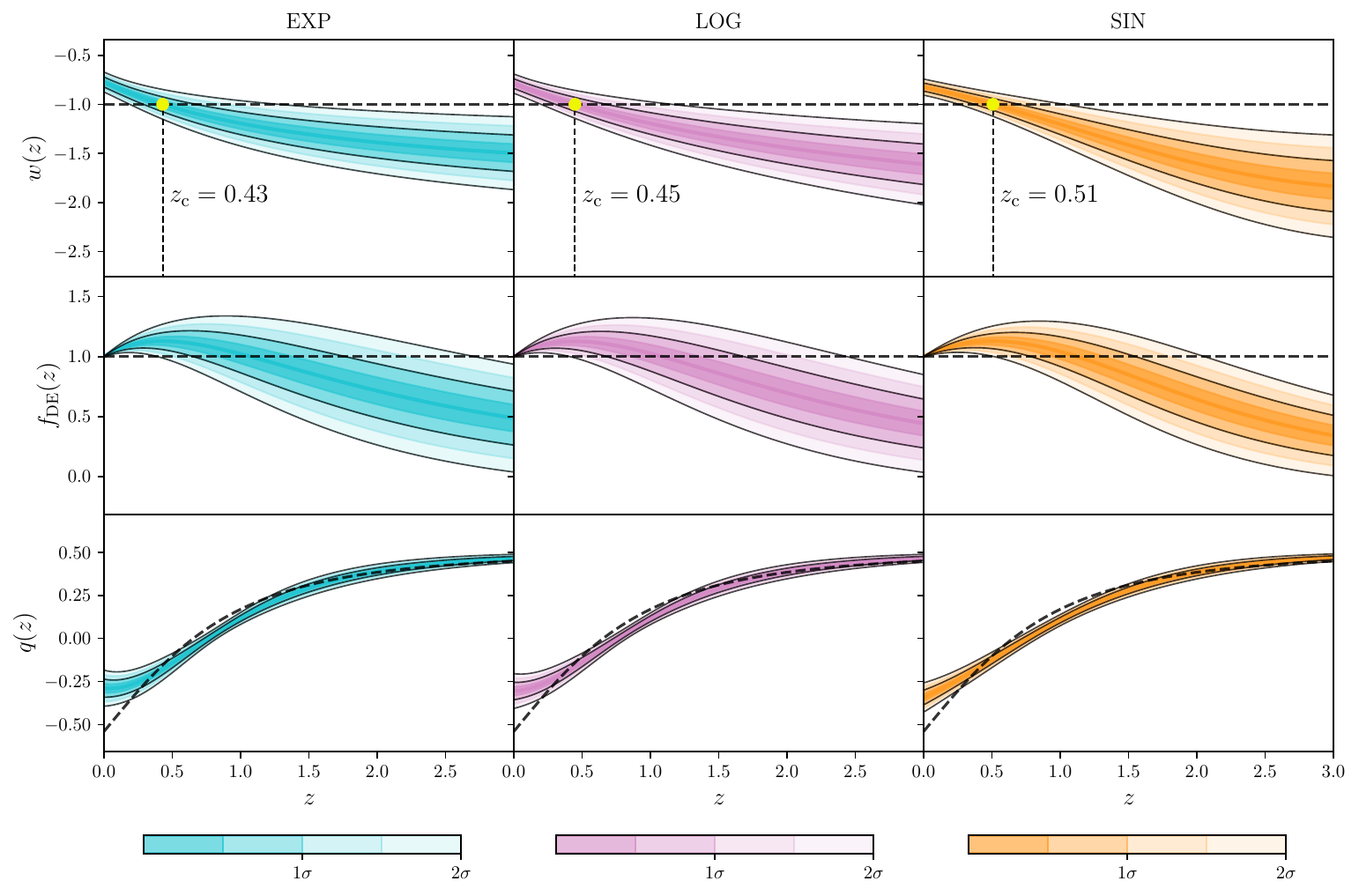}
\caption{Reconstruct redshift evolution of cosmological quantities: DE EoS $w(z)$, normalized DE density $f_{\mathrm{DE}}(z)$, and deceleration parameter $q(z)$ at $1\sigma$ and $2\sigma$ confidence levels in the CPL, JBP, BA, EXP, LOG, and SIN models, using CMB+DESI+DESY5 data. Here, the black dashed lines represent predictions from the $\Lambda$CDM model, and $z_{\rm c}$ represent the redshift of the $w=-1$ crossing.}
\label{fig3}
\end{figure*}

\begin{figure*}[htbp]
\includegraphics[scale=0.6]{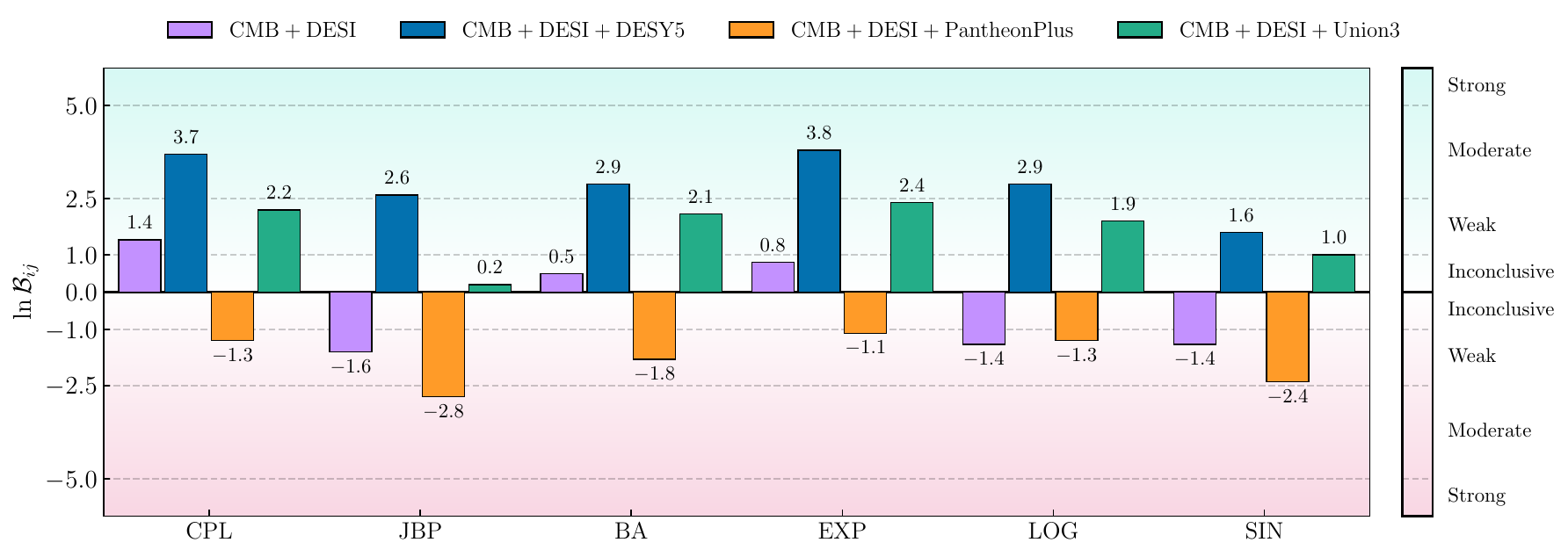}
\centering
\caption{\label{fig4} Comparison of the Bayesian evidence for the DDE models and the $\Lambda$CDM model. The Bayes factor $\ln \mathcal{B}_{ij}$ (where $i$ = DDE, $j$ = $\Lambda$CDM) and its strength according to the Jeffreys scale are used to assess the preference between models, where a positive value indicates a preference for the DDE model.}
\end{figure*}

In this section, we shall report the constraint results of the cosmological parameters. We consider the $\Lambda$CDM, CPL, JBP, BA, EXP, LOG, and SIN models to perform a cosmological analysis using current observational data, including the DESI DR2, ACT, SPT, Planck, DESY5, PantheonPlus, and Union3 data. We show the $1\sigma$ and $2\sigma$ posterior distribution contours for various cosmological parameters in these DDE models in Fig.~\ref{fig1}. The $1\sigma$ errors for the marginalized parameter constraints are summarized in Table~\ref{tab2}. We clearly present the evidence of each model under different datasets, as shown in Fig.~\ref{fig2}. In addition, we reconstruct the redshift evolution of cosmological quantities at the $1\sigma$ and $2\sigma$ confidence levels for these DDE models using the CMB+DESI+DESY5 data, as shown in Fig.~\ref{fig3}. Finally, we compare $\ln \mathcal{B}_{ij}$ of the DDE models relative to the $\Lambda$CDM model using the current observational data, as shown in Fig.~\ref{fig4}.

Similar to the DESI collaboration \cite{DESI:2024mwx,DESI:2025zgx}, we test the DDE models against four different data combinations involving DESI BAO measurements. In Fig.~\ref{fig1}, we present the constraint results in the $w_0$-$w_a$ plane for the CPL, JBP, BA, EXP, LOG, and SIN models based on CMB+DESI, CMB+DESI+DESY5, CMB+DESI+PantheonPlus, and CMB+DESI+Union3 data. For clarity, we have added two black lines ($w_0=-1$ and $w_0+w_a=-1$) to divide the parameter space into four distinct regions, corresponding to phantom, quintessence, Quintom-A, and Quintom-B. First, we discuss the polynomial form of the parameterization, as shown in the upper panel of Fig.~\ref{fig1}. For the CPL model, when using the CMB+DESI data, we find $w_0 = -0.430 \pm 0.210$, showing a $2.7\sigma$ deviation from $w_0 = -1$, while $w_a = -1.72 \pm 0.57$ provides hints of the evolution of the DE EoS parameter towards the Quintom-B regime. When the SN data are included, we observe a significant refinement of the constraints on the parameter space, reducing the error bars on the DE parameters by up to a factor of 3. In particular, for the CMB+DESI+DESY5, which yields $w_0 = -0.749 \pm 0.057$, we observe a $4.4\sigma$ deviation of $w_0$ from $-1$. Similarly, $w_a = -0.88^{+0.23}_{-0.19}$ is found to be significantly non-zero at the $3.8\sigma$ level, providing compelling evidence for DDE, with the corresponding parameter space clearly situated in the Quintom-B region. For the JBP model, unlike the CPL case, we find that the overall constraints become weaker and the central values further shift from $w_0=-1$ and $w_a=0$. When using the CMB+DESI+DESY5 data, we obtain $w_0 = -0.649^{+0.081}_{-0.073}$ and $w_a = -1.99\pm 0.45$, confirming that the evidence for DDE becomes much more pronounced compared to CPL, reaching a statistical significance of $\sim 4.3\sigma$ level. Moreover, in this scenario we obtain a lower value of $H_0 = 66.74 \pm 0.56~{\rm km}~{\rm s}^{-1}~{\rm Mpc}^{-1}$, in contrast to $H_0 = 67.98 \pm 0.24~{\rm km}~{\rm s}^{-1}~{\rm Mpc}^{-1}$ in the $\Lambda$CDM model. Consequently, this preference for a quintom-like behavior with $w_0 > -1$ and $w_a < 0$ further exacerbates the $H_0$ tension, which is consistent with the DESI results. For the BA model, compared to the CPL and JBP models, the parameters are more tightly constrained. For CMB+DESI, we obtain $w_0 = -0.500 \pm 0.180$, which is approximately $2.7\sigma$ level away from $w_0 = -1$. Additionally, $w_a = -0.83^{+0.29}_{-0.25}$ is approximately $2.8\sigma$ different from $w_a = 0$, demonstrating a preference for DDE similar to CPL. For CMB+DESI+PantheonPlus and CMB+DESI+Union3, the constraint values for $w_0$ are $-0.856 \pm 0.048$ and $-0.714 \pm 0.073$, as well as $w_a$ values of $-0.32 \pm 0.09$ and $-0.52^{+0.13}_{-0.12}$ (both significantly different from zero at more than $3.5\sigma$ level). Specifically, in the case of CMB+DESI+DESY5, $w_0 = -0.785 \pm 0.047$ shifts further away from $-1$, and $w_a = -0.43^{+0.10}_{-0.09}$ is found to be non-zero at more than $4.2\sigma$ level, further confirming the evidence for DDE.

Next, we discuss three special-function parameterizations, as shown in the lower panel of Fig.~\ref{fig1}. For the EXP model, we observe that the constraints are similar to those of the BA model. Specifically, for CMB+DESI+DESY5, we find $w_0 = -0.772 \pm 0.051$, significantly different from $-1$, while $w_a = -0.65^{+0.17}_{-0.15}$ is nearly $3.8\sigma$ away from $w_a = 0$, further supporting the preference for DDE. For the LOG model, compared to other parameterizations, $w_0$ and $w_a$ exhibit a stronger anti-correlation and show a significant deviation from the $\Lambda$CDM model ($w_0 = -1, w_a = 0$). In the case of CMB+DESI, $w_0 = -0.590 \pm 0.160$ is confined to the Quintom-B region at more than $2\sigma$ level. For $w_a = -4.70^{+1.80}_{-1.50}$, although the central value deviates significantly from zero, the error is large, with a significance of only about $2.6\sigma$ level away from zero. When PantheonPlus data is added to CMB+DESI, we obtain $w_0 = -0.858^{+0.044}_{-0.050}$; although $w_0$ shifts towards $w_0 = -1$, the error bars are reduced by about a factor of 3. Therefore, $w_0$ is still preferred to lie in the Quintom-B region, with $w_0 = -1$ excluded at a significance of approximately $2.8\sigma$. Similarly, the result for $w_a = -2.12^{+0.71}_{-0.61}$ confirms the overall preference for DDE at about $2.9\sigma$. When using CMB+DESI+DESY5 and CMB+DESI+Union3, the constraint values for $w_0$ are $-0.788 \pm 0.049$ and $-0.720^{+0.070}_{-0.078}$, with $w_0$ remaining in the Quintom-B region for both cases. The constraint values for $w_a$ are $-2.83^{+0.72}_{-0.65}$ and $-3.48^{+0.96}_{-0.85}$, with $w_a$ found to be significantly different from zero, at approximately $3.9\sigma$ and $3.6\sigma$, respectively. For the SIN model, we observe similar constraints to those of the CPL model, except in the case of CMB+DESI. In the CMB+DESI case, we obtain $w_0 = -0.700^{+0.120}_{-0.140}$; compared to the CPL model, $w_0$ shifts towards $w_0 = -1$ at only the $2.1\sigma$ level. When SN data (PantheonPlus, DESY5, and Union3) are added to CMB+DESI, the constraint precision is only slightly improved compared to the CPL model. For example, for CMB+DESI+DESY5, the constraints are $w_0 = -0.824 \pm 0.041$ with a significance of approximately $4.2\sigma$ and $w_a = -0.98^{+0.27}_{-0.24}$ which is found to be significantly different from zero, close to $3.8\sigma$ level. 

To more intuitively represent the evidence for DDE in the CPL, JBP, BA, EXP, LOG, and SIN models across different data combinations, we present these main results in Fig.~\ref{fig2}. Here, we primarily assess the statistical significance of the preference for DDE across different data combinations based on $\Delta \chi^2_\mathrm{MAP} \equiv \chi^2_{\rm DDE}-\chi^2_{\Lambda\rm CDM}$. Since the $\Lambda$CDM model is a special case of the DDE model, $\Delta\chi^2_\mathrm{MAP}$ is expected to follow a $\chi^2$ distribution with two degrees of freedom ($w_0$ and $w_a$), assuming the null hypothesis (the $\Lambda$CDM model) is true, and that the errors are Gaussian and appropriately estimated. To interpret $\Delta\chi^2_\mathrm{MAP}$ in more familiar terms, we refer to the corresponding frequentist significance $N\sigma$ for a one-dimensional Gaussian distribution, given by
\begin{equation}
    \mathrm{CDF}_{\chi^2}\left(\Delta\chi^2_\mathrm{MAP} |\, 2\,\mathrm{dof}\right) = \frac{1}{\sqrt{2\pi}}\int_{-N}^{N} e^{-x^2/2} \, {\rm d}x,
\end{equation}
where the left-hand side represents the cumulative distribution function of $\chi^2$.

When considering the CMB+DESI data, the evidence supporting DDE is similar across these models, approximately $3\sigma$. When adding PantheonPlus to CMB+DESI, the evidence for DDE weakens in some cases, with $2.3\sigma$ and $2.9\sigma$ for the CPL and JBP models, respectively. When replacing DESY5 with PantheonPlus, we find that the evidence for DDE strengthens, similar to the DESI study. We can clearly observe that, in all cases, the evidence for DDE is most significant for CMB+DESI+DESY5, ranging from approximately $3.8\sigma$ to $4.2\sigma$. Specifically, for the CPL, JBP, and BA models, the evidence for DDE reaches $3.9\sigma$, $4.1\sigma$, and $4.2\sigma$, respectively. When including Union3, the evidence for DDE lies between that of PantheonPlus and DESY5 in most cases. Interestingly, in most cases the BA model reports stronger evidence for DDE than all other models.

To better understand the DDE behaviour, we reconstruct the redshift evolution of cosmological quantities, including the DE EoS $w(z)$, normalized DE density $f_{\mathrm{DE}}(z)$, and deceleration parameter $q(z)$, based on CMB+DESI+DESY5 data, at the $1\sigma$ and $2\sigma$ confidence levels, as shown in Fig.~\ref{fig3}. The black dashed lines represents the trajectory corresponding to the $\Lambda$CDM model. We have provided $w(z)$ and $f_{\mathrm{DE}}(z)$ in Sec.~\ref{sec2.1}, and the deceleration parameter $q(z)$ is written as
\begin{equation}\label{eq:q(z)}
q(z)\equiv{-\frac{\ddot{a}a}{\dot{a}^2}}=\frac{{\rm d}\ln H}{{\rm d}\ln(1+z)}-1.
\end{equation}
This function acts as a sensitive probe for new physics, as it is solely dependent on the ``shape'' of the (normalized) expansion history. In the CPL model, significant dynamical evolution is exhibited, with $w(z)$ crossing $-1$ at $z_{\rm c} \simeq 0.40$. The CPL model provides greater flexibility in the EoS, as evidenced by the broader distribution of $w(z)$ at earlier times. For the JBP model, the evolution of the DE EoS presents a more complex phenomenology, which is the main reason for the larger uncertainties in the constraints on the DE EoS parameters. We observe that in the JBP parameterization, the EoS evolves across $w = -1$ twice, a result of its quadratic form in the scale factor. The first crossing occurs at approximately $z > 3$, followed by a decline towards a minimum value of $w(z)$ around $z \sim 1$, after which it rises again, leading to a second crossing from phantom to quintessence at $z_{\rm c} \simeq 0.30$. For the BA model, the most notable difference occurs before $z > 1$, where it remains in the phantom regime, stabilizing on a distinctive, nearly flat trend with $w(z)$ does not trend towards very negative values at $z \gtrsim 1$. At low redshifts, we find that it behaves similarly to the CPL model, with $w(z)$ crossing $-1$ at $z_{\rm c} \simeq 0.42$. For the EXP, LOG, and SIN models, the evolution of $w(z)$ follows a trend similar to that of the CPL model, with the only difference being that $w(z)$ crosses $-1$ earlier than in the CPL model, at $z_{\rm c} \simeq 0.43$, $z_{\rm c} \simeq 0.45$, and $z_{\rm c} \simeq 0.51$, respectively. By reconstructing $f_{\mathrm{DE}}(z)$, we find that these models exhibit a trend that varies with redshift, further indicating insights into the evolution of DE over time. As the redshift decreases, $f_{\mathrm{DE}}(z)$ gradually increases, approaching $f_{\mathrm{DE}}(z) = 1$ around $z \sim 1$. In the regime $z<1$ we find that $f_{\rm DE}(z)$ increases with redshift until it reaches a maximum and then decreases, with $f_{\rm DE}(z=0)=1$. Notably, the peak redshift at which $f_{\rm DE}$ attains its maximum coincides with the redshift $z_{\rm c}$ of the $w=-1$ crossing. This agreement is a direct consequence of the integral relation between $w(z)$ and $f_{\rm DE}(z)$, as shown in Eq.~(\ref{eq2}). Additionally, the reconstructed $q(z)$ suggests that these DDE models are largely similar, with the onset of cosmic acceleration ($q < 0$) occurring earlier than predicted by the $\Lambda$CDM model ($z \simeq 0.8$), and a slowdown in cosmic acceleration at recent times ($z \simeq 0.5$).

Finally, we apply the Bayesian evidence selection criterion to identify the preferred DDE models over the $\Lambda$CDM model based on the current observational data. To compute the Bayesian evidence for the models, we use the publicly available code {\tt MCEvidence}\footnote{\url{https://github.com/yabebalFantaye/MCEvidence}.}
 \cite{Heavens:2017afc, Heavens:2017hkr}. The Bayesian evidence $Z$ is defined as
\begin{equation}
Z = \int_{\Omega} P(D|\bm{\theta},M)P(\bm{\theta}|M)P(M){\rm d}\bm{\theta},
\label{eq: lnZ}
\end{equation}
where $P(D|\bm{\theta},M)$ is the likelihood of the observational data $D$ given the parameters $\bm{\theta}$ and the cosmological model $M$, $P(\bm{\theta}|M)$ is the prior probability of $\bm{\theta}$ given $M$, and $P(M)$ is the prior of $M$. We compute the Bayes factor in logarithmic space, defined as $\ln \mathcal{B}_{ij} = \ln Z_i - \ln Z_j$, where $Z_i$ and $Z_j$ represent the Bayesian evidence for the two models.

The strength of model preference is commonly assessed using the Jeffreys scale \citep{Kass:1995loi,Trotta:2008qt}. According to this scale, when $\left|\ln \mathcal{B}_{ij}\right| < 1$, the evidence is considered inconclusive; a value of $1 \leq \left|\ln \mathcal{B}_{ij}\right| < 2.5$ indicates weak evidence; $2.5 \leq \left|\ln \mathcal{B}_{ij}\right| < 5$ corresponds to moderate evidence; $5 \leq \left|\ln \mathcal{B}_{ij}\right| < 10$ is strong evidence; and if $\left|\ln \mathcal{B}_{ij}\right| \geq 10$, the evidence is considered decisive. 

In Fig.~\ref{fig4}, we show the Bayes factors $\ln \mathcal{B}_{ij}$ for the DDE models relative to the $\Lambda$CDM model, based on the current observational data. Here, $i$ denotes the DDE model, and $j$ denotes the $\Lambda$CDM model. It is important to emphasize that a positive value indicates a preference for the DDE models over the $\Lambda$CDM model. We find that the Bayes factor values for all DDE models are positive when using the CMB+DESI+DESY5 and CMB+DESI+Union3 data. Specifically, $\ln \mathcal{B}_{ij} = 3.7$, $2.9$, and $3.8$ for the CPL, BA, and EXP models using CMB+DESI+DESY5 data, respectively, indicating moderate evidence in favor of these models relative to the $\Lambda$CDM model. By contrast, for CMB+DESI+PantheonPlus, the Bayes factor values are negative for all DDE models. For example, the value of $\ln \mathcal{B}_{ij} = -2.8$ for the JBP model indicates moderate evidence disfavoring this model relative to the $\Lambda$CDM model. For CMB+DESI, there is only weak evidence favoring the CPL model relative to the $\Lambda$CDM model. Overall, the current observational data favor the CPL, BA, and EXP models relative to the $\Lambda$CDM model, especially the CPL model in most data combination cases.

Note that the apparent deviation of the DESI data from the $\Lambda$CDM model remains an open question. Beyond the DDE scenarios emphasized in this paper, the same deviation could in principle be explained by other new-physics mechanisms. For example, considering an interaction between DE and dark matter~\cite{Li:2024qso,Pan:2025qwy,Giare:2024smz,Li:2025owk}, dark matter with a non-zero EoS parameter~\cite{Li:2025dwz,Kumar:2025etf,Yang:2025ume,Wang:2025zri}, or modifications of gravity involving non-minimal coupling~\cite{Wolf:2025jed,Ye:2024ywg,Wang:2025znm} can likewise account for such a departure and be preferred by current observations (including DESI) relative to $\Lambda$CDM. This paper focuses on a comparative analysis of DDE models using the latest observational data\footnote{For earlier important comparative studies of DE models, see, e.g., Refs.~\cite{Li:2009bn,Li:2010xjz,Xu:2016grp,Guo:2018ans}.} and does not attempt a comprehensive comparison with those alternative explanations; such a systematic study is left for future investigation.

\section{Conclusion}\label{sec4}

The DESI collaboration has utilized DR2 BAO measurements, and when combined with CMB data and SN data, this suggests a $2.8\sigma$--$4.2\sigma$ preference for DDE \cite{DESI:2025zgx}. This finding poses a significant challenge to the $\Lambda$CDM interpretation of the current accelerated expansion of the universe, suggesting instead a DDE component, whose EoS was in the phantom regime in the past and has transitioned to quintessence regime today. Therefore, in this context, it is crucial to consider the latest and most precise cosmological data across representative parameterized models to test the robustness of the DDE results. 

In this work, our aim is to conduct a comprehensive analysis of DDE models using the BAO data from DESI DR2, CMB data from ACT, SPT, and Planck, as well as SN data from DESY5, PantheonPlus, and Union3. We consider six DDE models, including CPL, JBP, BA, EXP, LOG, and SIN. We explore the dynamical evolution of these models, as well as the extent to which the DDE model is supported by the current observational data.

Our overall analysis indicates that, in most cases of the parameterization or data combination, the evidence for a deviation from $\Lambda$CDM ($w_0 = -1,\; w_a = 0$) is significant, showing a clear preference for DDE in the Quintom-B regime ($w_0 > -1,\; w_0+w_a < -1$). Specifically, when using the CMB+DESI+DESY5 dataset the evidence for DDE across all models is at the $3.8-4.2\sigma$ level. In particular, for the BA model we obtain $w_0 = -0.785 \pm 0.047$ and $w_a = -0.43^{+0.10}_{-0.09}$, which significantly deviate from the $\Lambda$CDM values and provide evidence for DDE at $4.2\sigma$ level. By reconstructing relevant cosmological quantities for DDE, we find that the EoS in all DDE models consistently evolves from $w(z)<-1$ at early times to $w(z)>-1$ at late times, with the crossing redshift in the range $z_c \simeq 0.30-0.51$. Meanwhile, reconstructions of $f_{\rm DE}(z)$ and $q(z)$ likewise show clear departures from $\Lambda$CDM, reinforcing the case for DDE. Moreover, the Bayesian evidence analysis indicates that the CPL, BA and EXP models are moderately favored relative to $\Lambda$CDM based on the CMB+DESI+DESY5 data.

In summary, it is essential to perform decisive tests of DE and its possible deviations from the $\Lambda$CDM model with a combination of high-precision and complementary probes. The forthcoming full DESI data, along with the ongoing supernova measurements from the ZTF survey~\cite{Rigault:2024kzb} and Large Synoptic Survey Telescope~\cite{LSST:2008ijt}, will extend the Hubble diagram to very low redshifts, improving constraints on the DE EoS parameters. Meanwhile, data from Euclid~\cite{Euclid:2024yrr} will serve as an important cross-check for the DESI findings, helping to assess the impact of potential systematic errors, while the CMB-S4 experiment~\cite{CMB-S4:2016ple} will further tighten constraints on early-universe parameters, breaking degeneracies with late-time observables.

\section*{Acknowledgments}
We thank Yichao Li, Peng-Ju Wu, and Yi-Min Zhang for their helpful discussions. This work was supported by the National Natural Science Foundation of China (Grants Nos. 12533001, 12575049, and 12473001), the National SKA Program of China (Grants Nos. 2022SKA0110200 and 2022SKA0110203), the China Manned Space Program (Grant No. CMS-CSST-2025-A02), and the National 111 Project (Grant No. B16009). 

\bibliography{main}

\end{document}